# A radical approach to promote multiferroic coupling in double perovskites

M.P. Singh[1*], K.D. Truong[1], P. Fournier[1], P. Rauwel[2], E. Rauwel[3], L.P. Carignan[4], and D. Ménard[4]

[1] Département de Physique and RQMP, Université de Sherbrooke, Sherbrooke (QC), J1K 2R1 Canada

[2] Department of Ceramics and Glass Engineering, CICECO, Universidade de Aveiro, Aveiro P-3810-193, Portugal

[3] Department of Chemistry, CICECO, Universidade de Aveiro, Aveiro P-3810-193, Portugal

[4] Département de Génie Physique and RQMP, École Polytechnique, Montréal (QC), H3C 3A7, Canada

## Abstract

Double perovskites provide a unique opportunity to induce and control multiferroic behaviors in oxide systems. The appealing possibility to design materials with a strong coupling between the magnetization and the polarization fields may be achieved in this family since these magnetic insulators can present structural self-ordering in the appropriate growth conditions. We have studied the functional properties of $La_2CoMnO_6$ and $Bi_2CoMnO_6$ epitaxial thin films grown by pulsed laser deposition. Cation-ordered $La_2CoMnO_6$ films display a magnetic Curie temperature of 250 K while cation-disordered $Bi_2CoMnO_6$ films present ferromagnetism up to ~ 800 K. Such high transition temperature for magnetic ordering can be further tuned by varying the strain in the films indicating an important contribution from the structural characteristics of the materials. Our approach might be generalized for other oxide systems. At this end, our results are compared with other multiferroic systems. The roles of various cations, their arrangements and structural effects are further discussed.

*Corresponding author,* **Email:** mangala.singh@usherbrooke.ca

## 1. Introduction

Multiferroics form a surprising class of materials displaying simultaneously magnetism, ferroelectricity, and/or ferroelasticity in a single phase [1-3]. The coexistence of various electronic order parameters that can interact in a given system opens new opportunities to design unconventional spintronic devices and also poses new challenges for physics, chemistry, and technology [1-3]. However, their applicability depends on the nature of the actual coupling between the electronic, magnetic, and lattice order parameters. In this context, there are various stumbling blocks. The most challenging ones are how we can couple directly their molecular magnetic and electric fields responsible for ferromagnetism and ferroelectricity, and control eventually this coupling? Moreover, if this intimate coupling is achieved, how important is the material's response to an external applied field? Answers to these questions are hence crucial for understanding the microscopic mechanism behind the ferroic coupling and for their applicability in real devices.

In this context, double perovskites $A_2B'B''O_6$ (where B' has partially filled $e_g$-orbitals while B'' has empty $e_g$-orbitals or *vice-versa*) provide a unique opportunity to promote strong multiferroic behaviours because they are ferromagnetic insulators governed by the 180° superexchange process [4] with a potential to generate a polar character. In fact, the polar nature can be obtained if one can promote different chemical valencies for the B' and B'' cations in a self-ordered phase with B' and B" alternating along specific crystallographic directions (i.e. with a B'- O – B" – O – B' sequence). Of course, such effect is absent in a disordered phase. For example, it has recently been shown that the ordered $La_2NiMnO_6$ have a $Ni^{2+}/Mn^{4+}$ cationic configuration leading to large dielectric constant [5] whereas the disordered phase has a $Ni^{3+}/Mn^{3+}$ configuration [5-7]. An alternative path to induce a polar character in double perovskites may be achieved by choosing an appropriate A-site cation (*e.g.*, Bi, Pb) that may induce a long-range molecular electric field in the system without the requirement of B'/B'' self-ordering. For example, the complete substitution of $La^{3+}$ by $Bi^{3+}$ can induce a long-range in-built molecular electric field in $A_2CoMnO_6$ double perovskites due to the contribution of Bi $6s^2$ lone pair electrons similar to that observed in $BiMnO_3$ [8]. From this perspective, comparing La$_2$CoMnO$_6$ and Bi$_2$CoMnO$_6$ properties in thin films becomes an interesting benchmark to understand the important ingredients required to promote useful multiferroic properties.

There are several reports on the structural and magnetic properties of bulk La$_2$CoMnO$_6$ double perovskites while very little information can be found on bulk Bi$_2$CoMnO$_6$ (BCMO) [4-7, 9-10]. Nevertheless, the available data have clearly demonstrated that the cation-disordered materials exhibit multiple magnetic transitions while the ordered ones are characterized by a single ferromagnetic-to-paramagnetic transition (FM-T$_c$). For example, La$_2$CoMnO$_6$ in its cation-ordered phase displays a unique FM-T$_c$ around 245 K, while disordered samples present also a second magnetic transition around 145 K [4-7, 9-10]. From a structural point of view, it has been also established that a cation-ordered La-based double perovskite displays monoclinic P2$_1$/n symmetry while the disordered system presents the orthorhombic Pnma symmetry [5-7]. The recent demonstration of magnetodielectric effects in the La-based self-ordered double perovskites in the vicinity of their ferromagnetic transition close to room temperature is a clear indication of their potential for novel multifunctional devices [6]. However, they still remain challenging materials and the observation of a coupling between the magnetization and the polarization needs to be correlated to microscopic models. Thus, a deeper understanding and a meticulous correlation between their physical and structural properties become crucial.

Unlike the study of the bulk phases, growth of double perovskite thin films is quite recent [11-18]. Obviously, these films are of great interest since they represent the first step toward their integration in multifunctional applications for spintronic devices. This has motivated us to compare the functional properties of LCMO and BCMO films. In this paper, we present our results on epitaxial LCMO and BCMO double perovskite thin films grown by pulsed-laser deposition. The impact of the growth parameters on the Co/Mn ordering is presented as well as the functional magnetic properties of the films. The results reveal that the ordered LCMO films display ferromagnetism up to about 250K while the disordered BCMO films remain ferromagnetic up to anomalously large temperature of 800 K. We present a possible scenario to explain the surprising enhancement of the FM-$T_c$ of disordered BCMO films with respect to LCMO. This mechanism is based on the increase of the superexchange strength arising from the overlap between the Bi-$6s^2$ lone pair electrons and the Co-O-Mn, Mn-O-Mn, and Co-O-Co electronic wave functions aided by the structural changes.

## 2. Experimental details

Epitaxial films of $La_2CoMnO_6$ and $Bi_2CoMnO_6$ films were grown in the temperature range of 600-850 °C under 10-800 mTorr $O_2$ pressure by ablating the respective targets using a KrF laser ($\lambda$ = 248 nm) [15, 18]. Polycrystalline stoichiometric LCMO and off-stoichiometric BCMO (with 20 wt % excess Bi) targets were synthesized by standard solid state chemistry route. Unlike LCMO, BCMO films could be grown only in a very limited parameter window due to the high vapour pressure of Bi and a resulting low melting temperature. Films were grown on (100) $SrTiO_3$ (STO), (100) $LaAlO_3$ (LAO) and (100) Nb:$SrTiO_3$.

The crystallinity and epitaxial nature of these films were examined in detail using x-ray diffraction (XRD) with Cu-K$\alpha$ radiation in the $\theta$-$2\theta$, rocking curve, and $\phi$-scan modes. Moreover, the microstructure, crystallinity, and film-substrate interface quality were studied using a transmission electron microscope Hitachi H-9000 NA operating at 200 kV and equipped with an energy dispersive spectrometer (EDS). Magnetic properties of the films in the range of 10-400 K were studied using a SQUID magnetometer from Quantum Design. High temperature magnetic properties are studied using a vibrating sample magnetometer. Temperature and magnetic field dependence of the dielectric properties of the films were measured in a Physical property measurement system (PPMS) from Quantum Design using on a capacitor structure realized by depositing simply indium dots directly onto the LCMO/Nb:$SrTiO_3$ films.

## 3. Results and discussions

XRD studies show that the LCMO films on STO (001) grow in the (001) directions while the BCMO films on STO (001) grow in the (111)-direction [15, 18]. Furthermore,

LCMO films are characterized by sharper rocking curve peaks than the BCMO films. For example, the LCMO films grown under optimized conditions exhibit about 0.2° full width at half maximum (FWHM) while it is about 0.7° for BCMO films. Compared to LCMO, the relatively large FWHM for BCMO films can be interpreted as the natural outcome of the large strain expected from the important mismatch between cubic STO (001) and monoclinic BCMO (111). Despite the apparent incompatibility between STO substrate and the monoclinic crystal structures of LCMO and BCMO, we find that both LCMO and BCMO are growing coherently on STO.

The coherent can be confirmed in Fig.1a showing the HRTEM image of the LCMO/STO. The sharpness of the LCMO/STO interface can be directly inferred from the high-resolution transmission electron microscopy (HRTEM) images and the film is free from extended defects. Furthermore, LCMO films are also displaying a bi-domain microstructure growth, with domains rotated by 90° from each other. These resulting domains have the same growth directions but a different (in-plane) epitaxial relationship with the substrate: their tilt angle lies in perpendicular directions as confirmed from their respective electron diffraction patterns (not shown here). These patterns also display superlattice peaks confirming the presence of long-range Co/Mn ordering in LCMO. Similarly, the HRTEM image of BCMO/STO (Fig. 1b) shows a very sharp film-substrate interface. The detailed study shows however the absence of long-range Co/Mn ordering in this system (no detectable superlattice peaks).

The M-H loops measured at 10K on these films (Fig 2) show that LCMO has a saturation magnetisation close to 6 $\mu_B$/f.u. while it is lower in BCMO films (roughly 3$\mu_B$/f.u.). In the ordered $Co^{2+}$/$Mn^{4+}$ cationic configurations, the expected theoretical value of saturation magnetisation is 6$\mu_B$/f.u. [4, 5]. This indicates that LCMO has a well-ordered long-range $Co^{2+}$/$Mn^{4+}$ structure whereas BCMO films are disordered. In other words, the BCMO films possess a large proportion of the $Co^{3+}$/$Mn^{3+}$ disordered phase which is the leading factor for their low value of saturation magnetization. To verify further the long-range ordering in LCMO films, we measured their M-T curves under 0.05 kOe (inset of Fig 2a). One can observe a single magnetic transition temperature of about 250 K. It is consistent with the observation from their bulk counterparts and corroborates our structural findings [5]. Furthermore, we can also notice in Fig. 2b that the BCMO films are characterized by a 150 Oe coercivity, while LCMO films exhibit about 2.5 kOe coercivity. Interestingly, the LCMO films display double-hysteresis loops arising from the presence of a bi-domain structure [15]. It is a natural outcome of the bi-domain structural growth observed in the HRTEM images (Fig. 1a). The domains with their easy axis along the direction of the applied magnetic field can be easily magnetized and thus displaying a low coercive field and a well-defined square-shaped loop. On the contrary, the remaining domains require a large magnetic field to reach saturation or switch polarisation direction and hence would display a large coercivity and therefore a relatively large hysteresis area. Thus, the magnetic properties of these films are consistent with their microstructural properties.

Correlating the growth parameters (mostly the growth temperature and the $O_2$ pressure) and the structural and magnetic properties, we can draw a phase stabilization diagram of LCMO and BCMO as shown in Figure 3. Also the usual conditions used to grow the doped $LaMnO_3$ (*i.e.,* manganite) films are also included for comparison. Unlike the LCMO films, BCMO films can only be grown in a very narrow growth window. This is due to BCMO's low melting temperature and high vapour pressure of Bi. Furthermore, it is important to note that both materials possess the same B-site magnetic lattice (*i.e.,* Co/Mn). In the case of LCMO, long-range Co/Mn cations ordering can only be obtained in a very limited growth conditions. This requires high growth temperature as well as high $O_2$ pressure, which is far above the typical growth conditions used to deposit the BCMO films. It is also far above the classical growth parameters used to grow the manganite films by pulsed laser deposition. It is important to note that the films grown out of this high temperature/oxygen pressure zone present a large proportion of the disordered $Mn^{3+}/Co^{3+}$ phase (observed as two transitions in the M-T curves). This also explains why we obtained disordered BCMO films, a result also consistent with the literature on double perovskite films [11-18].

The M-T curve (inset of Fig. 2b) measured on a BCMO film in the range of 10-400 K shows a transition around 180 K, which is consistent with the other disordered double perovskites. However, the magnetization has only dropped by about 15% at 400 K from its 10K value illustrating that the main ferromagnetic transition has not yet been crossed. The presence of substantial M-H loops up to 400K with saturation magnetization of ~ 2.5$\mu_B$/f.u. confirms that the BCMO films are still ferromagnetic [18]. This demonstrates that the BCMO has a FM-$T_c$ above 400 K unlike LCMO which is ferromagnetic up to 250 K. High-temperature M-T curves (Fig. 4a), measured under 0.5 kOe, show that BCMO has a ferromagnetic-to-paramagnetic transition around 780 K. Thus the observed FM-$T_c$ value of BCMO films is far above the expected value of 100 K of bulk BCMO [9, 10].

By choosing the appropriate substrates (Fig 4a), it is also possible to alter the FM-$T_c$ value of BCMO thin films in the range of 750-800 K [18]. It is important to note that both BCMO and LCMO possess the same magnetic octahedra, *viz,* $CoO_6$ and $MnO_6$. The difference between the LCMO and BCMO systems is that BCMO has a long-range molecular electric field promoted by Bi $6s^2$ lone pair electrons while such effect is absent in LCMO.

As pointed out earlier, the magnetic properties of BCMO and LCMO are determined by the 180°-superexchange process [4]. In this mechanism, the amplitude of the spin transfer integral is determined by the Co-O-Mn bond-length and bond angle. Unlike LCMO, BCMO system exhibits a large degree of crystal distortion arising due to the Bi $6s^2$ lone pair electron leading to ferroelectricity [9, 10]. This may lead to significant changes in bond lengths and bond angles. However, it would be insufficient to provoke such a large ferromagnetic transition temperature. This requires an alternative superexchange mechanism to induce such a large transition temperature.

The ferromagnetic 180°-superexchange interaction can be also governed by polarization effects. In the present case, the large anisotropic overlap of these Bi $6s^2$ electrons with the surrounding hybridized Co-O-Mn, Co-O-Co, Mn-O-Mn orbitals can facilitate its coupling to the long-range spin order through the minimization of the magnetoelastic energy. This polarization effect does not only allow the direct coupling between the magnetization and the polarization fields, but it can also enhance drastically the ferromagnetic superexchange interactions leading to a ferromagnetic transition temperature close to its ferroelectric transition. It is important to note that such superexchange mechanism does not contribute in LCMO because the electronic states of $La^{3+}$ have little (or no) overlap with those of the surrounding 3d cations. The indirect impact of the crystal structure can be confirmed by the observed variation of the FM-$T_c$ from 700–800 K on the films grown on various substrates. These films show also different monoclinic β- tilt angle depending on the substrate used. A plot of the FM-$T_c$ as a function of this β-angle for our films (Fig. 4b) helps underline the relationship between the structure and the magnetic properties. For comparison, the value of FM-$T_c$ of bulk BCMO is also included in Fig. 4b. This data suggests that the structure, through the tilt angle, plays a crucial role in enhancing the overlap between the Bi $6s^2$ and 3d-O(2p) electronic wave functions. Thus, both correlation and polarization effects are likely to cooperate to enhance the strength of the ferromagnetic superexchange interactions giving rise to the anomalously large ferromagnetic transition temperature of our BCMO thin films. It is premature to predict that such enhancement in FM-$T_c$ from bulk BCMO data to strained thin films is taking place either by a crossover or through a sudden jump. This polarization-enhanced magnetic interaction may also be responsible for the large FM-$T_C$ in $Bi_2FeCrO_6$ films and $La_{0.7}Ca_{0.3}MnO_3/BaTiO_3$ superlattices [19, 20]. Nonetheless, these recent studies clearly points out that a direct coupling between the electric and magnetic molecular fields may induce peculiar magnetoelectric behaviours beyond our present understandings and requires further deeper studies.

To investigate the presence of a magnetoelectric coupling effect in LCMO films, we have studied the temperature dependence of their magnetodielectric behaviours [12]. To do so, we measured the temperature dependence of the dielectric constant at zero and 1 tesla magnetic fields. The magnetodielectric constant ($\varepsilon_{MD}$) is defined as [ε(10 kOe) - ε(0 Oe)]/ ε(0 Oe at 10K)). Fig 5 shows $\varepsilon_{MD}$ as a function of temperature. The dielectric constant of the ordered LCMO films is hardly varying with applied magnetic field in the paramagnetic phase, whereas a significant enhancement is observed in the ferromagnetic phase regime giving rise to a peak near FM-$T_c$. Similar magnetodielectric effects have been reported in $La_2NiMnO_6$ and $BiMnO_3$ ferromagnetic insulators [6, 12, 21-22]. The observed large magnetodielectric effect can be roughly explained using the phenomenological Landau theory giving $\varepsilon_{MD} = \gamma M^2$, where M is the magnetization, and $\gamma$ is the magnetoelectric coupling constant. The large dielectric constant ($\varepsilon \sim 250$) and the related magnetodielectric effect at FM-$T_c$ in LCMO result from the coupling of the local polarisation order parameter to the short wavelength magnetic fluctuations which are important only in the vicinity of the ferromagnetic-to-paramagnetic transition temperature [12, 21, 22]. This explains the significant variation in the magnetodielectric constant only around the FM-$T_c$. Further work is in progress to study the magnetoelectric properties of BCMO films.

## 4. Concluding remarks

In summary, we have studied the functional properties of thin films of a polar-ferromagnet $La_2CoMnO_6$ and a ferroelectric-ferromagnet $Bi_2CoMnO_6$. The growth parameters promoting the long-range Co/Mn structural ordering were obtained in unusual ranges when compared to the growth conditions of manganites. Our study shows that Co/Mn-ordered $La_2CoMnO_6$ display ferromagnetism up to 250 K while Co/Mn-disordered $Bi_2CoMnO_6$ display ferromagnetism for temperature as high as 800 K. This observed enhancement is explained based on a possible superexchange mechanism involving the direct interaction between the local magnetisation and the $Bi6s^2$ polarization aided by the crystal structure. Furthermore, we demonstrate a correlation between the magnetodielectric effect, magneto-elastic interactions, and magnetic properties of $Bi_2CoMnO_6$ and $La_2CoMnO_6$ films.

## Acknowledgements

We thank S. Pelletier and M. Castonguay for their technical support. This work was supported by *CIFAR*, *CFI*, *NSERC* (Canada), *FQRNT* (Québec) and the *Université de Sherbrooke*. Two of us (P Rauwel and E Rauwel) also acknowledge the financial support from FCT grant No. SFRH/BPD/36941/2007 and Marie Curie (MEIF-CT2006-041632).

**References:**

1. W. Prellier, M.P. Singh, and P. Murugavel, J. Phys.: Cond. Maters. **17**, (2005) R803.
2. N.A. Hill, J. Phys. Chem B. **104** (2000) 6694. M. Fiebig, J. Phys. D **38** (2005) R123.
3. R. Ramesh and N.A. Spladin, Nature Materials **6** (2007) 21. S.W. Cheong and M. Mostovoy, Nature Materials **6** (2007) 13.
4. J. B. Goodenough, *Magnetism and the chemical bond,* chapter 3 (Inter Science Pub. NY, 1976).
5. R I. Dass, J. Q. Yan, and J.B. Goodenough. Phys. Rev. B **67** (2003) 014401; *ibid*, **68** (2003) 064415.
6. N. S. Rogado, J. Li, A.W. Sleight, and M.A. Subramanian, Adv. Mater. **17** (2005) 2225.
7. C. L Bull, D. Gleeson and K. S. Knight, J. Phys. Condens. Mat. **15** (2003) 4927.
8. N.A. Hill and K.M. Rabe, Phys. Rev B **59** (1999) 8759.
9. M. Azuma, K. Takata, T. Saito, S. Ishiwata, Y. Shimakawa, and M. Takano, J. Am. Chem. Soc. **127** (2005) 8889.
10. K. Takata, M. Azuma, Y. Shimakawa, and M. Takano, J. Jpn. Soc. Powder Metall. **52** (2005) 913.
11. M.P. Singh, C. Grygiel, W.C. Sheets, Ph. Boullay, M. Hervieu, W. Prellier, B. Mercey, Ch. Simon, and B. Raveau, Appl. Phys. Lett. **91** (2007) 012503.
12. M.P. Singh, K.D. Truong, and P. Fournier, Appl. Phys. Lett. **91** (2007) 042504.
13. H. Guo, A. Gupta, T. G. Calvarese, and M.A. Subramanian, Appl. Phys. Lett. **89** (2006) 262503.
14. H. Guo, A. Gupta and J. Zhang, Appl. Phys. Lett. 91 (2007) 202509.
15. M. P. Singh, S. Charpentier, K. D. Truong, and P. Fournier, Appl. Phys. Lett. **90** (2007) 211915. M. P. Singh, K. D. Truong, J. Laverdière, S. Charpentier, S. Jandl, and P. Fournier, J. Appl. Phys. **103** (2008) 07E315.
16. M. N. Iliev, M. V. Abrashev, A. P. Litvinchuck, V.G. Hadjiev, H. Guo, A. Gupta, Phys. Rev. B. **75** (2007) 104118.
17. K.D. Truong, J. Laverdière, M.P. Singh, S. Jandl, and P. Fournier, Phys. Rev. B **76** (2007)132413.

18. M.P. Singh, K. D. Truong, P. Fournier, P. Rauwel, E. Rauwel, L. P. Carignan, and D. Ménard, Appl. Phys. Lett. **92** (2008) 112505.
19. R. Nechache, C. Harnagea, A. Pignolet, F. Normandin, T. Veres, L. P. Carignan, and D. Menard, Appl. Phys. Lett. **89** (2006) 102902.
20. M.P. Singh, W. Prellier, L. Mechin, and Ch. Simon, Thin Solid Films 515 (2007) 6526.
21. G. Lawes, A.P. Ramirez, C.M. Varma, M.A. Subramanian, Phys. Rev. Lett. **91**, (2003) 257208.
22. T. Kimura, S. Kawamoto, I. Yamada, M. Azuma, Y. Takano, and Y. Tokura, Phys. Rev. B **67** (2003) 180401.

**Figure Captions:**

**Figure 1:** Cross-sectional HRTEM images of **(a)** $La_2CoMnO_6$/$SrTiO_3$ (001) illustrating a sharp interface and the presence of bi-domains structure (red arrows) and **(b)** $Bi_2CoMnO_6$/$SrTiO_3$ (001) sharp interface.

**Figure 2:** In-plane M-H loops of **(a)** $La_2CoMnO_6$ and **(b)** $Bi_2CoMnO_6$ films measured at 10 K. The insets show their respective magnetization as a function of temperature.

**Figure 3:** Phase-stability diagram for the growth of $La_2CoMnO_6$ and $Bi_2CoMnO_6$. The usual growth conditions for manganites thin films are also presented. This diagram illustrates that relatively high temperatures and oxygen pressures are required to obtain the self-ordered Co/Mn sublattice in double perovskites.

**Figure 4:** (Color online) **(a)** Magnetization as a function of temperature for $Bi_2CoMnO_6$ films grown on different substrates. **(b)** Variation in FM-$T_C$ as a function of the $\beta$-angle of BCMO deduced from structural data. For sake of comparison the bulk BCMO value is also included. The dashed and solid lines present two different scenarios for the variation of FM-$T_c$ with the tilt angle.

**Figure 5:** Magnetodielectric constant of a $La_2CoMnO_6$ film as a function of temperature demonstrating a strong coupling between magnetic and polar order parameters.

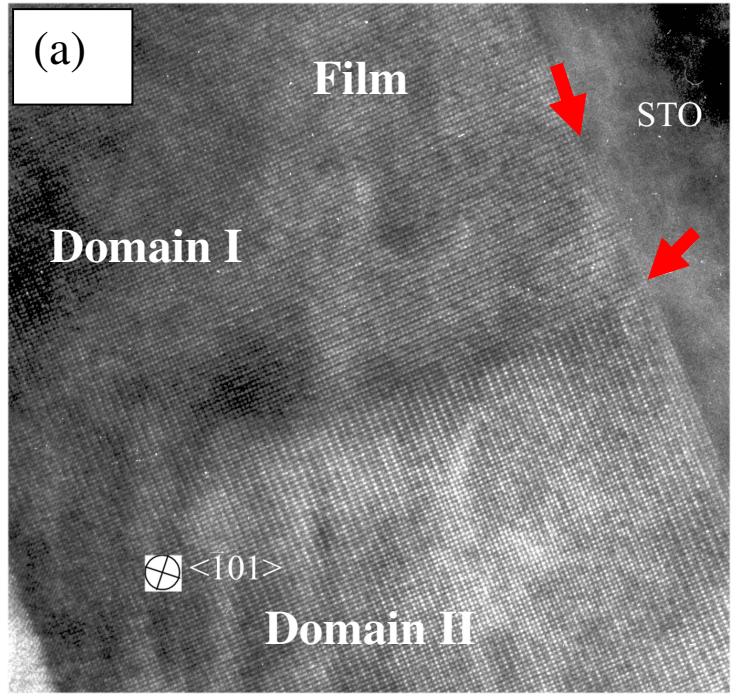

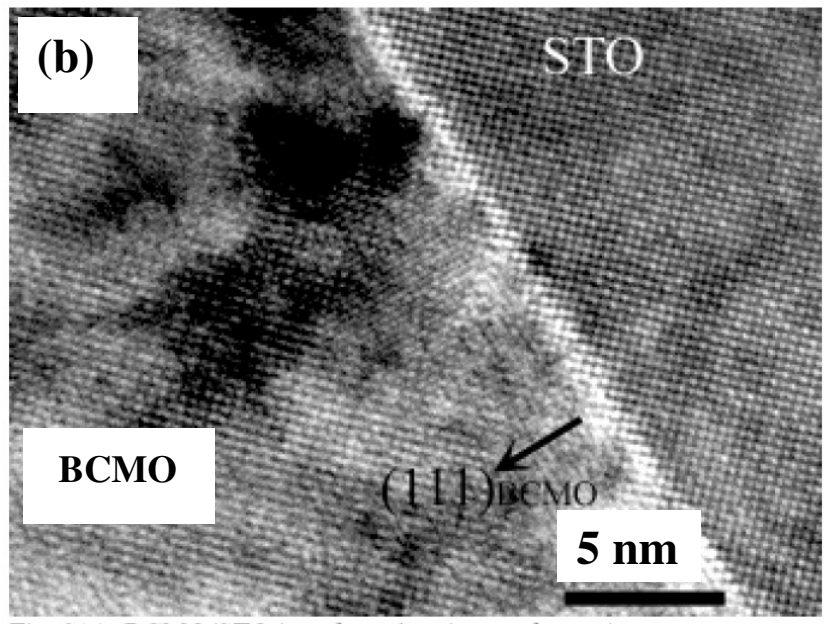

Figure 1

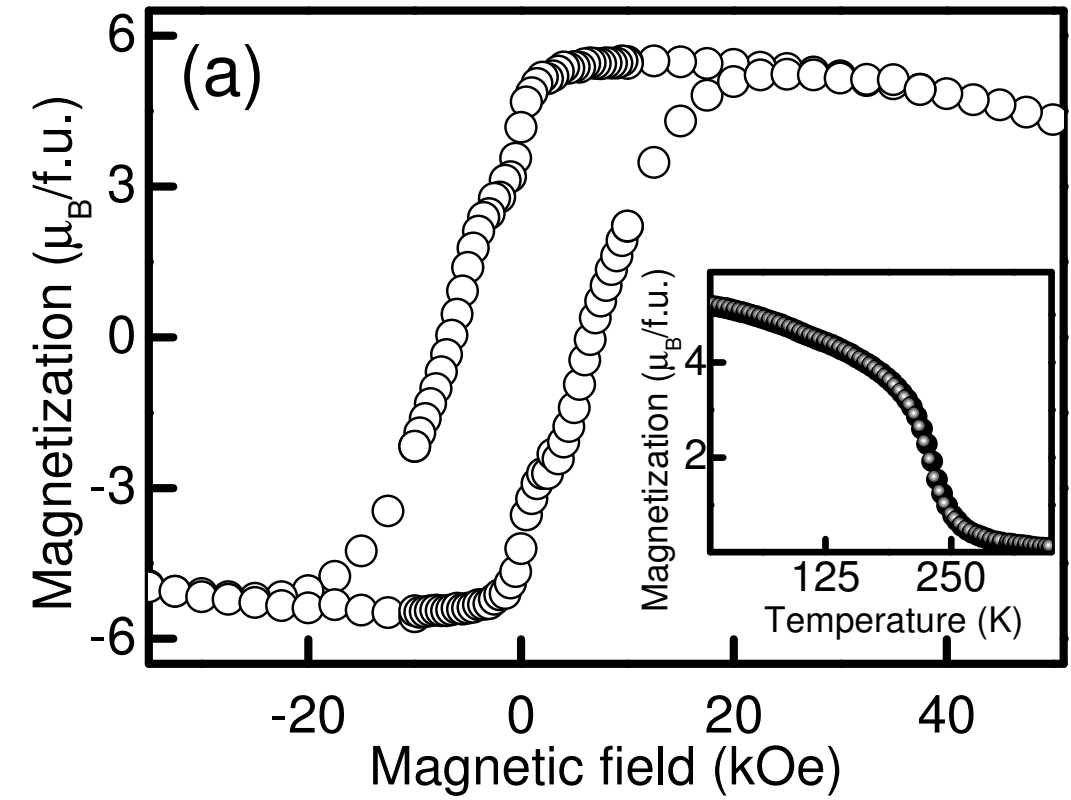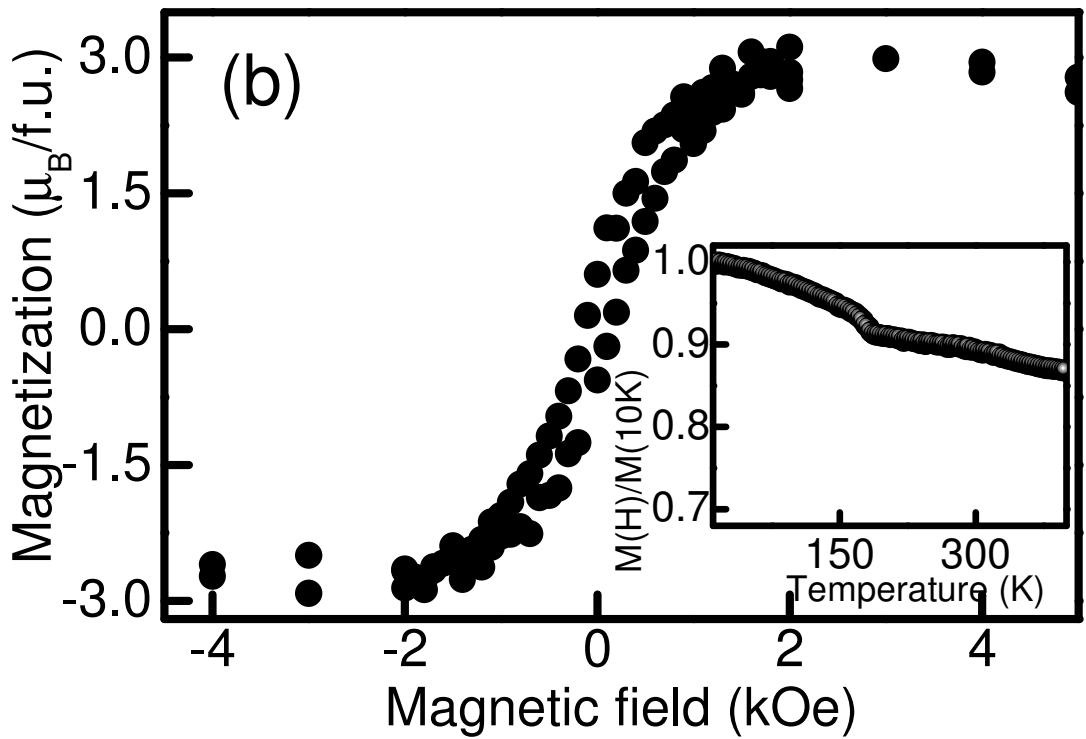

Figure 2

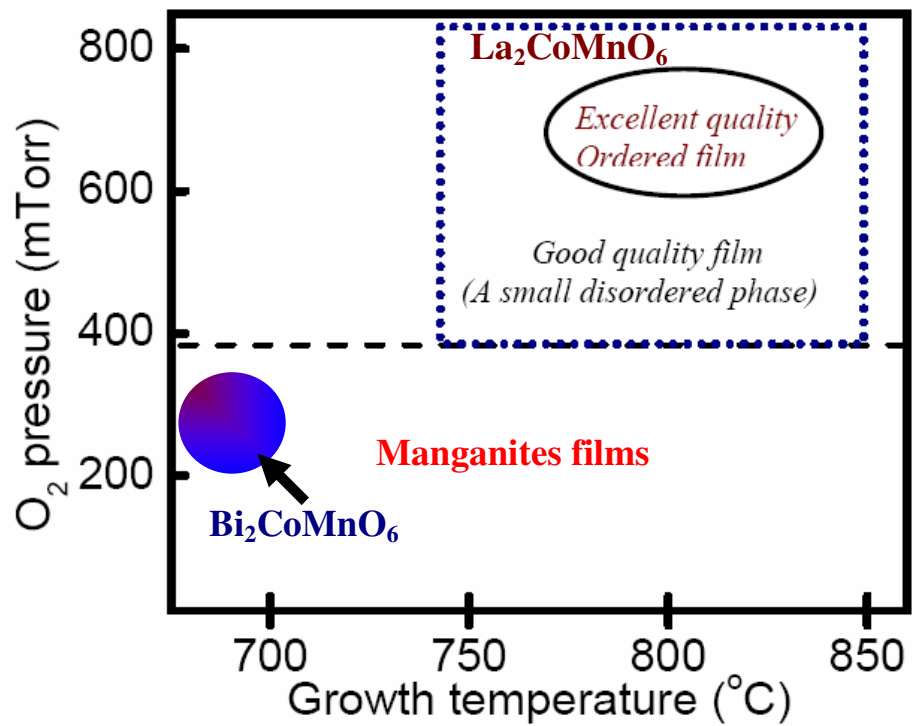

Figure 3

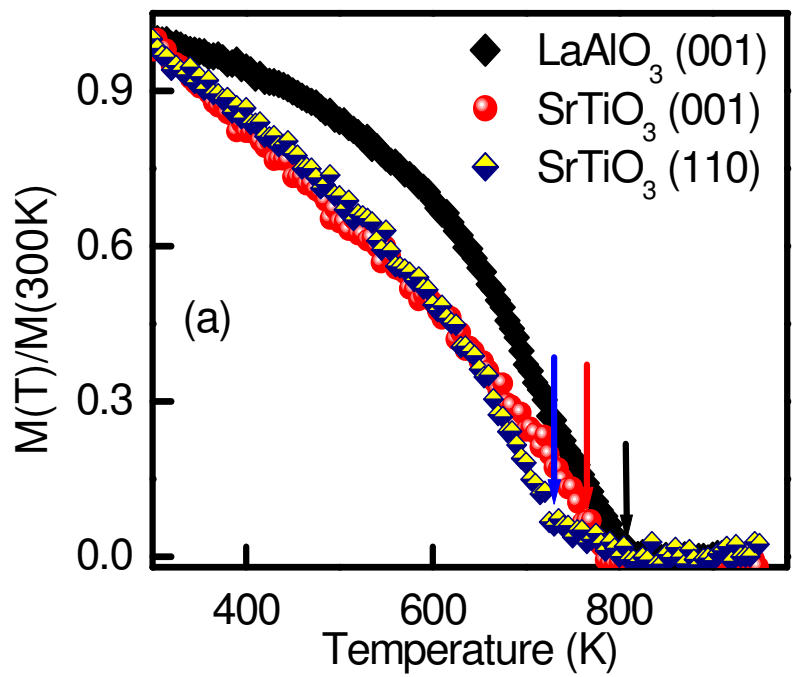

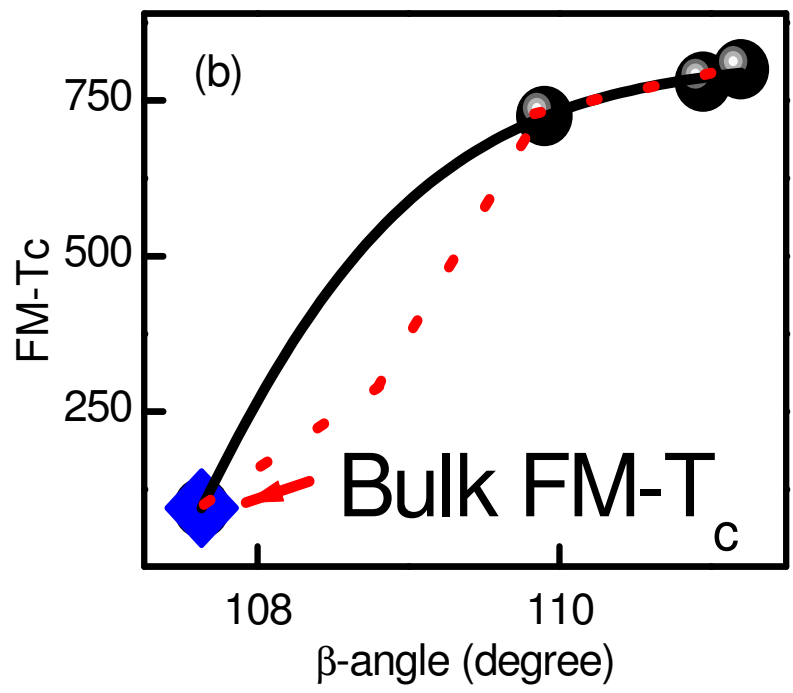

Figure 4

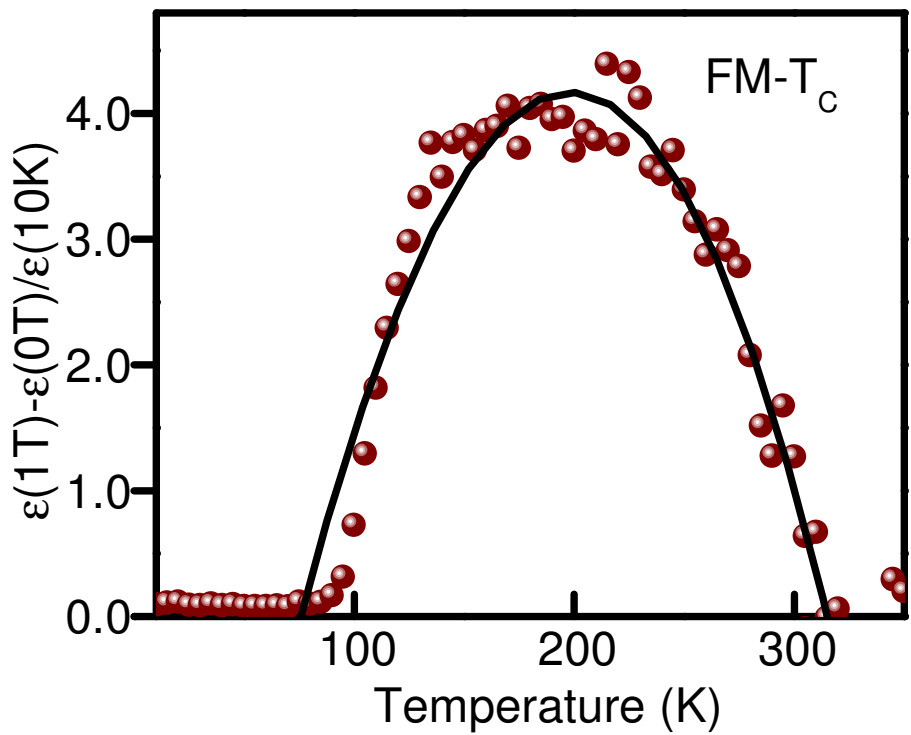

Figure 5